\documentclass[fleqn,10pt]{wlscirep}

\usepackage{dcolumn}       
\usepackage{bm}            
\usepackage{amsmath}       
\usepackage{graphicx}      
\usepackage{hyperref}      
\usepackage{amssymb}       
\usepackage{latexsym}      
\usepackage{epstopdf}     
\usepackage{color}
\usepackage{subfigure}

\newcommand{\bs}[1] {  \boldsymbol{#1}           }
\newcommand{\ff}[1] {  \mbox{\footnotesize{#1}}  }

\title{Applying machine learning techniques to predict the properties of energetic materials}

\author[1,*]{Daniel C.\ Elton}
\author[1]{Zois Boukouvalas}
\author[1]{Mark S.\ Butrico}
\author[1]{Mark D.\ Fuge}
\author[1,**]{Peter W.\ Chung}
\affil[1]{Department of Mechanical Engineering, University of Maryland, College Park, 20742, United States}

\begin{abstract}
We present a proof of concept that machine learning techniques can be used to predict the properties of CNOHF energetic molecules from their molecular structures. We focus on a small but diverse dataset consisting of 109 molecular structures spread across ten compound classes. Up until now, candidate molecules for energetic materials have been screened using predictions from expensive quantum simulations and thermochemical codes. We present a comprehensive comparison of machine learning models and several molecular featurization methods - sum over bonds, custom descriptors, Coulomb matrices, Bag of Bonds, and fingerprints. The best featurization was sum over bonds (bond counting), and the best model was kernel ridge regression. Despite having a small data set, we obtain acceptable errors and Pearson correlations for the prediction of detonation pressure, detonation velocity, explosive energy, heat of formation, density, and other properties out of sample. By including another dataset with $\approx 300$ additional molecules in our training we show how the error can be pushed lower, although the convergence with number of molecules is slow. Our work paves the way for future applications of machine learning in this domain, including automated lead generation and interpreting machine learning models to obtain novel chemical insights.
\end{abstract}
\begin{document}

\flushbottom
\maketitle
* delton@umd.edu ~~~ ** pchung15@umd.edu

\thispagestyle{empty}

\maketitle

\section{Introduction}
During the past few decades, enormous resources have been invested in research efforts to discover new energetic materials with improved performance, thermodynamic stability, and safety. A key goal of these efforts has been to find replacements for a handful of energetics which have been used almost exclusively in the world's arsenals since World War II - HMX, RDX, TNT, PETN, and TATB.\cite{council2004report} While hundreds of new energetic materials have been synthesized as a result of this research, many of which have remarkable properties, very few compounds have made it to industrial production. One exception is CL-20,\cite{Nielsen1998:11793,Viswanath2018} the synthesis of which came about as a result of development effort that lasted about 15 years.\cite{council2004report} After its initial synthesis, the transition of CL-20 to industrial production took another 15 years.\cite{council2004report} This time scale (20-40 years) from the initial start of a materials research effort until the successful application of a novel material is typical of what has been found in materials research more broadly. Currently, the development of new materials requires expensive and time consuming synthesis and characterization loops, with many synthesis experiments leading to dead ends and/or yielding little useful information. Therefore, computational screening and lead generation is critical to speeding up the pace of materials development. Traditionally screening has been done using either ad-hoc rules of thumb, which are usually limited in their domain of applicability, or by running large numbers of expensive quantum chemistry calculations which require significant supercomputing time. Machine learning (ML) from data holds the promise of allowing for rapid screening of materials at much lower computational cost. A properly trained ML model can make useful predictions about the properties of a candidate material in milliseconds rather than hours or days.\cite{Gilmer2017arxiv}

Recently, machine learning has been shown to accelerate the discovery of new materials for dielectric polymers,\cite{Mannodi-Kanakkithodi2016:20952} OLED displays,\cite{gomez-bombarelli2016:1120} and polymeric dispersants.\cite{Menon2017} In the realm of molecules, ML has been applied successfully to the prediction of atomization energies,\cite{Rupp2012:058301} bond energies,\cite{Kun2017:2689} dielectric breakdown strength in polymers,\cite{Pilania2013:2810} critical point properties of molecular liquids,\cite{Carande2015:1377} and exciton dynamics in photosynthetic complexes.\cite{Hase2016:5139} In the materials science realm, ML has recently yielded predictions for dielectric polymers,\cite{Pilania2013:2810,Mannodi-Kanakkithodi2016:20952
} superconducting materials,\cite{Stanev2017arxiv} nickel-based superalloys,\cite{Conduit2017:358} elpasolite crystals,\cite{Faber2016:135502} perovskites,\cite{Schmidt2017:5090} nanostructures,\cite{Ju2017:021024} Heusler alloys,\cite{Sanvitoe2017:1602241} and the thermodynamic stabilities of half-Heusler compounds.\cite{Fleur2017} In the pharmaceutical realm the use of ML has a longer history than in other fields of materials development, having first been used under the moniker of quantitative structure-property relationships (QSPR). It has been applied recently to predict properties of potential drug molecules such as rates of absorption, distribution, metabolism, and excretion (ADME),\cite{Maltarollo2015:259} toxicity,\cite{Mayr2016:80}  carcinogenicity,\cite{Zhang2017:2118} solubility, and binding affinity.\cite{Junshui2015:263}

As a general means of developing models for relationships that have no known analytic forms, machine learning holds great promise for broad application. However, the evidence to date suggest machine learning models require data of sufficient quality, quantity, and diversity which ostensibly limits the application to fields in which datasets are large, organized, and/or plentiful. Published studies to date use relatively large datasets, with $N = 10,000 - 100,000$ being typical for molecular studies and larger datasets appearing in the realm of materials. When sufficiently large datasets are available, very impressive results can be obtained. For example, recently it has been shown that with enough data ($N = 117,000$\cite{Faber2017:5255} or $N = 435,000$\cite{Ward2017:12}) machine learning can reproduce properties calculated from DFT with smaller deviations from DFT values than DFT's deviation from experiment.\cite{Faber2017:5255,Ward2017:12}

Compared to other areas of materials research, the application of ML methods to energetic materials has received relatively little attention likely due to the scarcity of quality energetics data. Thus energetics may be a suitable platform to consider the factors that effect machine learning performance when limited to small data. While machine learning has been applied to impact sensitivity,\cite{Rice2002:1770,Prana2012:169,Devinyak2014:194,XU2012:10} there is little or no previously published work applying ML to predict energetic properties such as explosive energy, detonation velocity, and detonation pressure. While there is previous work applying ML to heat of formation \cite{Fayet2011:2443,Turker2010:139} and detonation velocity \& pressure,\cite{Turker2011:7,Infante-Castillo2012:304686,Ravi2012:218,Zeman2007Book} the studies are restricted to energetic materials in narrow domains, such as series of molecules with a common backbone. Furthermore, the aforementioned studies have been limited to the use of hand picked descriptors combined with linear modeling.

Therefore, in this work we wish to challenge the assumption that large data sets are necessary for ML to be useful by doing the first comprehensive comparison of ML methods to energetics data. We do this using a dataset of 109 energetic compounds computed by Huang \& Massa.\cite{Huang_2013:197} While we later introduce additional data from Mathieu\cite{Mathieu2017:8191} for most of our work we restrict our study to the Huang \& Massa data to demonstrate for the first time how well different ML models \& featurizations work with small data. The Huang \& Massa data contains molecules from ten distinct compound classes and models trained on it should be relatively general in their applicability. The diverse nature of the Huang \& Massa data is important as ultimately we wish our models to be applicable to wide range of candidate energetic molecules.

To obtain the energetic properties in their dataset, Huang \& Massa calculated gas phase heats of formation using density functional theory calculations at the B3LYP/6-31G(d,p) level, and calculated the heat of sublimation using an empirical packing energy formula.\cite{Huang2011:33} These two properties were used to calculate the heat of formation of the solid $\Delta H_f^{\ff{solid}}$. They obtained densities primarily from experimental crystallographic databases. Huang \& Massa then used their calculated heats of formation and densities as inputs to a thermochemistry code which calculates energetic properties under the Chapman-Jouguet theory of detonation. They validated the accuracy of their method with experimental data.\cite{Huang2011:33} The result is a dataset with nine properties - density, heat of formation of the solid, explosive energy, shock velocity, particle velocity, sound velocity, detonation pressure, detonation temperature, and TNT equivalent per cubic centimeter. Several of these properties have significant correlations between them (a correlation matrix plot is given in Supplementary Fig. S1).

\section{Featurization methods}
In the case of small data featurization is more critical than selecting a model, as with larger data models can learn to extract complex and/or latent features from a general purpose (materials agnostic) featurization. Feature vectors must be of reasonable dimensionality $d << N$, where $d$ is the number of dimensions of the feature vector and $N$ is the number of molecules in the training set, to avoid the curse of dimensionality and the so-called ``peaking phenomena".\cite{Theodoridis2008book} 
Some featurizations make chemically useful information more transparent than others. More abstract general purpose featurizations, such as SMILES strings \& Coulomb matrices (discussed below) only excel with large data and deep learning models, which can learn to extract the useful information. With small data, great gains in accuracy can sometimes be gained by hand selecting features using chemical intuition and domain expertise. For example, the number of azide groups in a molecule is known to increase energetic performance while also making the energetic material more sensitive to shock. While the number of azide groups is implicitly contained in the SMILES string and Coulomb matrix for the molecule, ML models such as neural networks typically need a lot of data to learn how to extract that information. To ensure that azide groups are being used by the model to make predictions with small data, an explicit feature corresponding to the number of such groups can be put in the feature vector.

\subsection{Oxygen balance}
It is well known that the energy released during an explosion is largely due to the reaction of oxygen with the fuel atoms carbon and hydrogen. Based on this fact, Martin \& Yallop (1957) found a linear relationship between detonation velocity and a descriptor called oxygen balance.\cite{Martin1958:257} It can be defined either of two ways:
\begin{equation}\label{Obalance}
\hbox{OB}_{1600} \equiv \frac{1600}{m_{\ff{mol}}}( n_{\ff{O}} - 2 n_{\ff{C}} - n_{\ff{H}}/2)
\hspace{1cm}
\hbox{OB}_{100} \equiv \frac{100}{n_{\ff{atoms}}}( n_{\ff{O}} - 2 n_{\ff{C}} - n_{\ff{H}}/2)
\end{equation}
Here $n_{\ff{C}}$, $n_{\ff{H}}$, and $n_{\ff{O}}$ are the number of carbons, hydrogens, and oxygens respectively, $m_{\ff{mol}}$ is the molecular weight, and $n_{\ff{atoms}}$ is the number of atoms. An oxygen balance close to zero is sometimes used as requirement in determining if material may be useful as a novel energetic.\cite{klapotke2017book} While it is easy to calculate and provides a useful rule of thumb, oxygen balance has limitations, which will become clear when compared to more sophisticated featurizations. One limitation of oxygen balance is that it neglects the large variation in bond strengths found in different molecules. It also neglects additional sources of energy released in an explosion, such as from nitrogen recombination (formation of N$_2$), halogen reactions, and the release of strain energy. Furthermore, oxygen balance is built on the assumption that oxygen reacts completely to form CO$_2$ and H$_2$O. More sophisticated calculations take into account the formation of CO (which commonly occurs at high temperatures or in low-oxygen compounds) as well as H$_2$ and trace amounts of unreacted O$_2$ and solid carbon which may appear. Predicting the proportion of such products requires complex thermochemical codes.

\subsection{Custom descriptor set}
By a ``descriptor" we mean any function that maps a molecule to a scalar value. There are many types of descriptors, ranging from simple atom counts to complex quantum mechanical descriptors that describe electron charge distributions and other subtle properties.
There are many different algorithms for generating a descriptor set of a specified size from a large pool. Since a brute force combinatorial search for the best set is often prohibitively expensive computationally, usually approximate methods are employed such as statistical tests, greedy forward elimination, greedy backwards elimination, genetic algorithms, etc. For instance, Fayet et al.\ used Student's $t$-test to rank and filter descriptors from a pool of 300 possible ones.\cite{Fayet2011:2443} A pitfall common to the $t$-test and other statistical tests is that two descriptors that have low ranks on their own may be very useful when combined (eg. through multiplication, addition, subtraction, division). Given many of the difficulties in descriptor set selection, Guyon and Elisseeff recommend incorporating physical intuition and domain knowledge whenever possible.\cite{Guyon2003:1157}

We chose to design our descriptor set based on physical intuition and computational efficiency (we ignore descriptors which require physics computations). The first descriptor we chose was oxygen balance. Next we included the nitrogen/carbon ratio ($n_N/n_C$), a well known predictor of energetic performance.\cite{Politzer2014book} Substituting nitrogens for carbon generally increases performance, since N$=$N bonds yield a larger heat of formation/enthalpy change during detonation compared to C-N and C$=$N bonds.\cite{Politzer2014book} In addition to raw nitrogen content, the way the nitrogen is incorporated into the molecule is important. For instance, the substitution of N in place of C-H has the extra effect of increasing crystal density. Nitrogen catenation, both in N$=$N and N-N$\equiv$N (azide) is known to greatly increase performance but also increase sensitivity. On the other hand, nitrogen in the form of amino groups (NH$_2$) is known to decrease performance but also decrease sensitivity.\cite{Politzer2014book} The way oxygen is incorporated into a molecule is similarly important - oxygen that is contained in nitro groups (NO$_2$) release much more energy during detonation than oxygen that is already bonded to a fuel atom. Martin \& Yallop (1958)  distinguish four oxygen types that are present in energetic materials.\cite{Martin1958:257} Based on all of this, we distinguished different types of nitrogen, oxygen, and flourine based on their bond configurations:

\begin{tabular}{ll}
     N-N-O$_2$ (nitrogen nitro group) & C=N-O (fulminate group)   \\
     C-N-O$_2$ (carbon nitro group)   & C-N=N (azo group)  \\
     O-N-O$_2$ (oxygen nitro group)   & C-N-H$_2$ (amine group)\\
     O-N=O  (nitrite group)           & C-N(-O)-C (N-oxide nitrogen)\\
     C=N-F  (nitrogen-flourine group) & C-F (carbon-flourine group) \\
     C-O-H (hydroxyl oxygen)          & N=O (nitrate or nitrite oxygen),\\
     N-O-C (N-oxide oxygen)           & C=O (keton/carboxyl oxygen)  \\
\end{tabular}

We also included raw counts of carbon and nitrogen, hydrogen, and fluorine. We tested using ratios instead of raw counts ($n_x / n_{\ff{total}}$) but found this did not improve the performance of the featurization. All together, our custom descriptor set feature vector is:
\begin{equation}
    \begin{aligned}
    \bs{x}_{\ff{CDS}} = &
    [ \hbox{OB}_{100},
    n_{\ff{N}}/n_{\ff{C}},
    n_{\ff{NNO2}},
    n_{\ff{CNO2}},
    n_{\ff{ONO2}},
    n_{\ff{ONO}},
    n_{\ff{CNO}},
    n_{\ff{CNN}},
    n_{\ff{NNN}},
    n_{\ff{CNH2}},\\
    & n_{\ff{CN(O)C}},
    n_{\ff{CNF}},
    n_{\ff{CF}},
    n_{\ff{C}},
    n_{\ff{N}},
    n_{\ff{NO}},
    n_{\ff{COH}},
    n_{\ff{NOC}},
    n_{\ff{CO}},
    n_{\ff{H}},
    n_{\ff{F}},
    ]
     \end{aligned}
\end{equation}

\subsection{Sum over bonds}
Based on the intuition that almost all of the latent heat energy is stored in chemical bonds, we introduce a bond counts feature vector. The bond counts vectors are generated by first enumerating all of the bond types in the dataset and then counting how many of each bond are present in each molecule. There are 20 bond types in the Huang \& Massa dataset:\\
\indent N-O, N:O, N-N, N=O, N=N, N:N, N\#N, C-N, C-C, C-H, C:N, C:C, C-F, C-O, C=O, C=N, C=C, H-O, H-N, F-N \\
We use the SMARTS nomenclature for bond primitives (`-' for single bond, `=' for double bond, `\#' for triple bond, and `:' for aromatic bond). Following Hansen, et al.\ \cite{Hansen2015:2326} we call the resulting feature vector ``sum over bonds". 

\begin{figure}[ht]
	\centering
	\includegraphics[width = .45\textwidth]{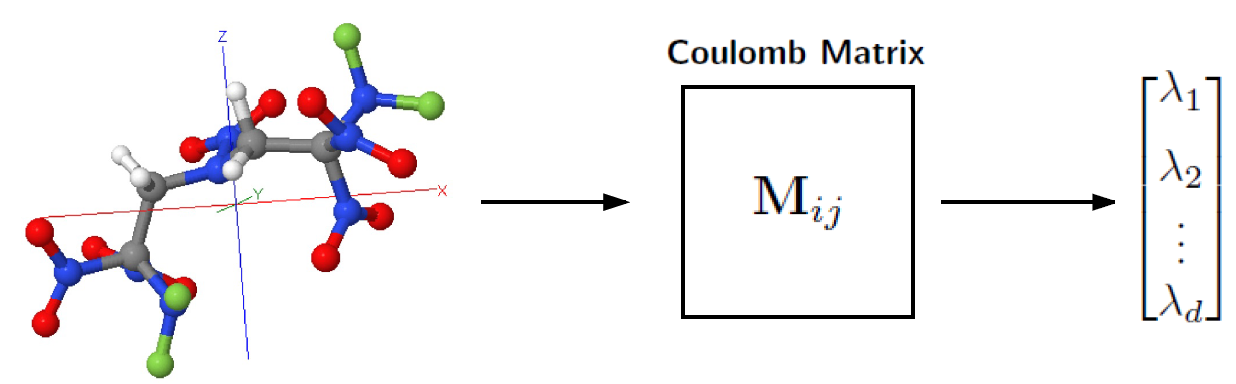}
	\caption{Transformation of (x,y,z) coordinates and nuclear charges to the Coulomb matrix eigenvalue spectra representation of a molecule.}
	\label{Spectra}
\end{figure}
\subsection{Coulomb matrices}
An alternative way of representing a molecule is by the Coulomb matrix featurization introduced by Rupp et al.\cite{Rupp2012:058301,Rupp:NIPS2012_4830}. The Coulomb matrix finds its inspiration in the fact that (in principle) molecular properties can be calculated from the Schr\"{o}dinger equation, which takes the Hamiltonian operator as its input. While extensions of the Coulomb matrix for crystals have been developed,\cite{Faber2015:1094} it is designed for treating gas phase molecules. The Hamiltonian operator for an isolated molecule can be uniquely specified by the nuclear coordinates $\bs{R}_i$ and nuclear charges $Z_i$. Likewise, the Coulomb matrix is completely specified by $\lbrace \bs{R}_i , Z_i \rbrace $. The entries of the Coulomb matrix ${\bf M}$ for a given molecule are computed as:
\begin{equation}\label{Coulob}
{\bf M}_{ij} = \begin{cases}
0.5Z_i^{2.4}, & i=j \\
\displaystyle \frac{Z_i Z_j}{|{\bf R}_i -  {\bf R}_j|}, & i\neq j
\end{cases},
\end{equation}
The diagonal elements of ${\bf M}$ correspond to a polynomial fit of the potential energies of isolated atoms, while the off-diagonals elements correspond to the energy of Coulombic repulsion between different pairs of nuclei in the molecule. It is clear by construction that ${\bf M}$ is invariant under translations and rotations of the molecule. However, Coulomb matrices are not invariant under random permutations of the atom's indices. To avoid this issue, one can use the eigenvalue spectrum of the Coulomb matrix since the eigenvalues of a matrix are invariant under permutation of columns or rows. In this approach, the Coulomb matrix is replaced by a feature vector of the eigenvalues, $\left< \lambda_1,\cdots,\lambda_d \right>$, sorted in a descending order, as illustrated in fig.\ \ref{Spectra}.

To obtain fixed length feature vectors, the size is of the vectors is set at the number of atoms in the largest molecule in the dataset $d = d_{\ff{max}}$ and the feature vectors for molecules with number of atoms less than $d_{\ff{max}}$ are padded with zeros. Using the eigenvalues implies a loss of information from the full matrix. Given this fact, we also compare with the ``raw" Coulomb matrix. Since it is symmetric, we take only the elements from the diagonal and upper triangular part of the matrix and then put them into a feature vector (which we call ``Coulomb matrices as vec"). The ``raw" Coulomb matrix is quite a bit different than other feature vectors as the physical meaning of each specific element in the feature vector differs between molecules. This appears to be problematic, especially for kernel ridge regression (KRR) and the other variants on linear regression, which treat each element separately and for which it is usually assumed that each element has the same meaning in every sample.

\subsection{Bag of bonds}
The Bag of Bonds featurization was introduced by Hansen et al.\ in 2015.\cite{Hansen2015:2326} It is inspired by the ``bag of words'' featurization used in natural language processing. In bag of words, a body of text is featurized into a histogram vector where each element, called a ``bag", counts the number of times a particular word appears. Bag of bonds follows a similar approach by having ``bags" that correspond to different types of bonds (such as C-O, C-H, etc). Bonds are distinguished by the atoms involved and the order of the bond (single, double, triple). However Bag of Bonds differs from Bag of Words in several crucial ways. First, each ``bag" is actually a vector where each element is computed as $Z_i Z_j / |\bf{R}_i - \bs{R}_j|$. The bag vectors between molecules are enforced to have a fixed length by padding them with zeros. The entries in each bag vector are sorted by magnitude from highest to lowest to ensure a unique representation. Finally, all of the bag vectors are concatenated into a final feature vector.

\subsection{Fingerprinting}
\begin{figure}[ht]
 \centering
 \includegraphics[width=9cm]{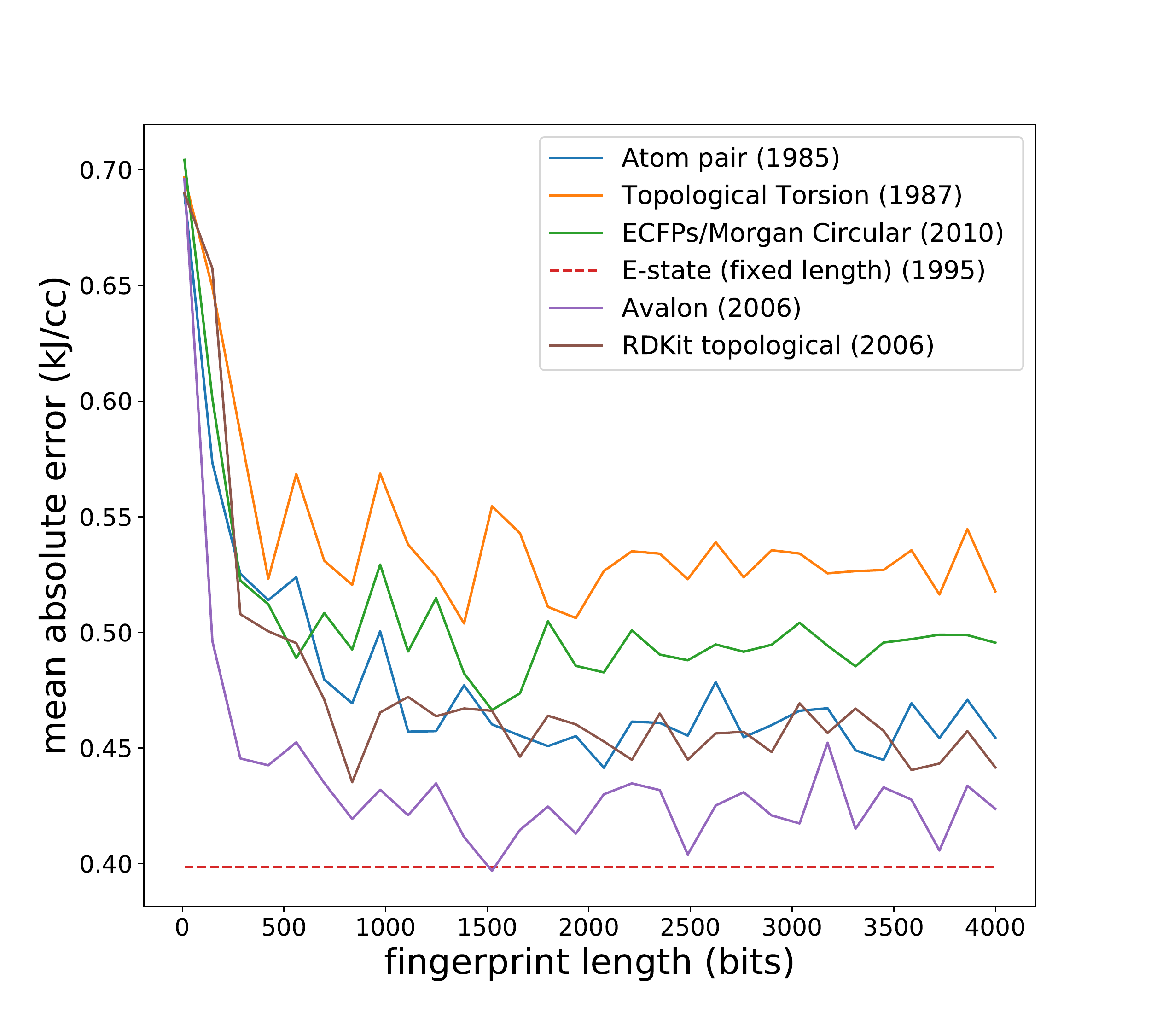}
  \caption{Mean errors in explosive energy obtained with different fingerprinting methods at different fingerprint lengths, using Bayesian ridge regression and leave-5-out cross validation. The E-state fingerprint has a fixed length and therefore appears as a flat line.}
   \label{fingerprintcomparison}
\end{figure}
Molecular fingerprints were originally created to solve the problem of identifying isomers,\cite{Morgan1965:107} and later found to be useful for rapid substructure searching and the calculation of molecular similarity in large molecular databases. In the past two decades, fingerprints have been used as an alternative to descriptors for QSPR studies. Fingerprinting algorithms transform the molecular graph into a vector populated with bits or integers. In this work we compare several fingerprints found in RDKit, a popular cheminformatics package --
Atom-Pair,\cite{Carhart1985:64} Topological Torsion,\cite{Ramaswamy1987:82} Extended Connectivity Fingerprints (ECFPs),\cite{Rogers2010:742} E-state fingerprints,\cite{Kier1995:1039} Avalon fingerprints,\cite{Gedeck2006:1924} RDKit graph fingerprints,\cite{rdkit} ErG fingerprints,\cite{Stiefl2006:208} and physiochemical property fingerprints.\cite{Kearsley1996:118}.

A few general features of fingerprints can be noted. First of all, all the fingerprinting algorithms start with atom level descriptors, each of which encode atom-specific information into an integer or real number. The simplest atom descriptor is the atomic number, but in most fingerprints this is augmented with additional information that describes the local environment of the atom. A second general point is that some fingerprinting algorithms can generate either integer vectors or bit vectors. In extended connectivity fingerprints (ECFPs) for instance, count information on the number of times a feature/fragment appears is lost when bit vectors are used. In RDKit, fingerprints return bit vectors by default. The length of bit vector based fingerprint representations can be tuned (to avoid the curse of dimensionality) through a process called folding.
Another point is that the fingerprints we study contain only 2D graph information, and do not contain information about 3D structure/conformation. Comparisons between 2D fingerprints and analogous 3D fingerprints have found that 3D fingerprints generally do not yield superior performance for similarity searching,\cite{Rhodes2006:615} or binding target prediction,\cite{Nettles2006:6802} although they are useful for certain pharmaceutical applications.\cite{Lowis1998:3}

\subsection{Other featurizations}
Two other featurizations that have been developed for molecules are smooth overlap of atomic positions (SOAP),\cite{Bartok2013:184115,Bartok2017scienceadvancespreprint} and Fourier series of atomic radial distribution functions.\cite{vonLilienfeld2015:1084} We choose not to investigate these featurizations due to their poorer performance in past comparisons. Another featurization strategy, the random walk graph kernel, may be promising for future exploration but is computationally expensive to properly implement.\cite{Ferre2017:114107} Very recently, several custom deep learning architectures with built-in featurization have been developed - custom graph convolutional fingerprints,\cite{Duvenaud2015:2224,Kearnes2016:595} deep tensor networks,\cite{Schutt2016:13890} message passing neural networks,\cite{Gilmer2017arxiv} and
hierarchically interacting particle neural nets.\cite{Lubbers2017arxiv} We attempted to train a graph convolutional fingerprint using the ``neural fingerprint" code of Duvenaud et al.\cite{Duvenaud2015:2224} but were not able achieve accuracy that was competitive with any of the other featurizations (for example the best we achieved for shock velocity was a MAE of $\approx 0.35$ km/s as opposed to 0.30 km/s for sum over bonds+KRR). Additional data and hyperparameter optimization would likely improve the competitiveness of custom convolutional fingerprinting.


\section{Results}
\begin{table*}
\centering
\begin{tabular}{c c c c c c c c c}
                   name          & MAE$_{\ff{train}}$   &  MAE$_{\ff{test}}$  & MAPE$_{\ff{test}}$ & R$^2_{\ff{train}}$ &  R$^2_{\ff{test}}$ &  $r_{\ff{train}}$ & $r_{\ff{test}}$          \\
\hline
           E-state + CDS + SoB &  0.244 &0.334 & 8.93 &     0.88 &  0.76 &  0.88 &  0.79  \\
                     CDS + SoB &  0.247 &0.335 & 9.32 &     0.88 &  0.75 &  0.88 &  0.79  \\
E-state + custom descriptor set &  0.224 &0.345 & 9.50 &     0.89 &  0.75 &  0.90 &  0.79  \\
                    SoB + OB100 &  0.256 &0.358 & 10.50 &     0.87 &  0.61 &  0.87 &  0.70  \\
          sum over bonds (SoB) &  0.280 &0.379 & 10.69 &     0.84 &  0.67 &  0.84 &  0.71  \\
             truncated E-state &  0.260 &0.414 & 12.65 &     0.85 &  0.66 &  0.85 &  0.70  \\
   custom descriptor set (CDS) & 0.398 &0.432 & 12.92 &     0.68 &  0.57 &  0.68 &  0.63  \\
            Bag of Bonds (BoB) &  0.213 &0.467 & 12.60 &     0.89 &  0.54 &  0.90 &  0.60  \\
       Oxygen balance$_{1600}$ &  0.419 &0.489 & 15.66 &     0.67 &  0.41 &  0.68 &  0.56  \\
           Summed Bag of Bonds &  0.262 &0.493 & 13.63 &     0.85 &  0.18 &  0.85 &  0.56  \\
    Coulomb matrix eigenvalues &  0.314 &0.536 & 15.73 &     0.81 &  0.37 &  0.82 &  0.48  \\
        Oxygen balance$_{100}$ &  0.444 &0.543 & 17.46 &     0.59 &  0.44 &  0.62 &  0.57  \\
       Coulomb matrices as vec &   0.395 &0.672 & 21.86 &     0.57 &.05 &  0.67 &  0.20  \\

\end{tabular}
\caption{\label{featurizationcomparison} Detailed comparison of 13 different featurization schemes for prediction of explosive energy with kernel ridge regression, ranked by MAE$_{\ff{test}}$. These variance in MAEs between folds was less than 0.01 in all cases. Hyperparameter optimization was used throughout with nested 5-fold cross validation. The metrics are averaged over 20 train-test sets using shuffle split with 80/20 splitting.}
\end{table*}

\subsection{Comparison of fingerprints}

Figure \ref{fingerprintcomparison} shows fingerprint mean absolute error of explosive energy in out-of-sample test data using kernel ridge regression vs.\ the fingerprint length in bits. The E-state fingerprint, which has a fixed length, performs the best, with the Avalon fingerprint nearly matching it. The Estate fingerprint is based on electrotopological state (E-state) indices,\cite{Kier1990:801} which encode information about associated functional group, graph topology, and the Kier-Hall electronegativity of each atom. E-state indices were previously found to be useful for predicting the sensitivity ($h_{50}$ values) of energetic materials with a neural network.\cite{Wang2009:155}
The E-state fingerprint differs from all of the other fingerprints we tested as it is fixed length, containing a vector with counts of 79 E-state atom types. Thus, technically it is more akin to a descriptor set than a fingerprint. We found that only 32 of the atom types were present in the energetic materials we studied (types of C,N,H,O,F), so we truncated it to a length of 32 (ie.\ threw out the elements that were always zero). The E-state fingerprint can also be calculated as a real-valued vector which sums the E-state indices for each atom, however we found the predictive performance was exactly the same as with the count vector version. It is not immediately clear why E-state performs better than the other fingerprints, but we hypothesize that it is due to the fact that the E-state atom types are differentiated by valence state and bonding pattern, including specifically whether a bond to a hydrogen is present. Thus a NH$_2$ nitrogen is considered a specific atom type. In the atom-pair, topological torsion, Morgan, and RDKit topological fingerprint the atom types are differentiated soley by atomic number, number of heavy atom neighbors, and number of pi electrons. The Avalon fingerprint, which does almost as well as E-state, likely does so because it contains a large amount of information including atom pairs, atom triplets, graph topology, and bonding configurations.

\subsection{Comparison of featurizations}
Table \ref{featurizationcomparison} shows a comparison of the featurization schemes discussed in the previous sections, using kernel ridge regression with hyperparameter optimization performed separately via grid search with cross validation for each featurization. Hyperparameter optimization was done for all hyperparameters - the regularization parameter $\alpha$, kernel width parameter and kernel type (Laplacian/L1 vs Gaussian/L2). Fairly dramatic differences are observed in the usefulness of different featurizations. The sum over bonds featurization always performs the best, but additional gains can be made by concatenating featurizations. The gains from concatenation are especially reflected in the correlation coefficient $r$, which increases from 0.65 to 0.78 after E-state and the custom descriptor set is concatenated with sum over bonds. As expected, the two different oxygen balance formulae perform nearly the same, and Coulomb matrix eigenvalues perform better than the raw Coulomb matrix. Summed bag of bonds (summing each bag vector) performs nearly as well as traditional bag of bonds, but with a much more efficient representation ($\bs{x} \in \Bbb{R}^{20}$ vs $\bs{x} \in \Bbb{R}^{2527}$).

\begin{table*}
\begin{tabular}{ccccccccccc}
 &  & \footnotesize{$\rho ,\frac{\hbox{g}}{\hbox{cc}}$ } & \footnotesize{$\Delta H_f^{\ff{s}} ,\frac{\hbox{kJ}}{\hbox{mol}}$ } & \footnotesize{$E_{\ff{e}} ,\frac{\hbox{kJ}}{\hbox{cc}}$ } & \footnotesize{$V_{\ff{s}} ,\frac{\hbox{km}}{\hbox{s}}$ } & \footnotesize{$V_{\ff{p}},\frac{\hbox{km}}{\hbox{s}}$ } & \footnotesize{$V_{\ff{snd}},\frac{\hbox{km}}{\hbox{s}}$ } & \footnotesize{$P$, GPa} & \footnotesize{$T$, K} & \footnotesize{$\frac{\hbox{TNT}_{\ff{equiv}}}{\hbox{cc}}$ } \\
\hline
KRR & Estate &  0.10 & 261.02 &  0.63 &  0.48 &  0.13 &  0.41 &  4.95 & 500.19 &  0.18\\
 & CDS &  0.08 & 198.81 &  0.50 &  0.44 &  0.11 &  0.37 &  3.07 & 462.63 &  0.17\\
 & SoB &  0.07 & 68.73 & 0.40 & \bf{ 0.31} & \bf{ 0.09} & \bf{ 0.25} & 2.90 & 331.36 & \bf{ 0.11}\\
 & CM eigs &  0.09 & 288.41 &  0.67 &  0.67 &  0.18 &  0.61 &  5.67 & 600.08 &  0.22\\
 & Bag of Bonds & \bf{ 0.06} & 166.66 &  0.47 &  0.33 &  0.11 & 0.29 &  3.38 & 478.93 &  0.18\\
 & \footnotesize{Estate+CDS+SoB} & \bf{ 0.06} & \bf{71.40} & \bf{ 0.36} &  0.32 &  0.10 &  0.29 &  2.76 & 359.66 &  0.13\\
Ridge & Estate &  0.09 & 269.11 &  0.58 &  0.57 &  0.14 &  0.45 &  4.71 & 491.21 &  0.19\\
 & CDS &  0.07 & 193.19 &  0.43 &  0.39 &  0.11 &  0.33 &  3.23 & 438.27 &  0.17\\
 & SoB & \bf{ 0.06} & 82.00 & 0.37 &  0.32 &  0.10 &  0.29 &  3.01 & \bf{327.43} & \bf{ 0.11}\\
 & CM eigs &  0.09 & 355.12 &  0.79 &  0.60 &  0.16 &  0.55 &  5.82 & 590.69 &  0.19\\
 & Bag of Bonds & \bf{0.06} & 163.76 &  0.48 &  0.32 &  0.11 &  0.31 &  3.37 & 472.93 &  0.19\\
 & \footnotesize{Estate+CDS+SoB} &  0.06 & 77.31 &  0.39 &  0.32 &  0.10 &  0.28 &  2.78 & 383.07 &  0.13\\
SVR & Estate &  0.09 & 207.78 &  0.60 &  0.45 &  0.13 &  0.35 &  4.41 & 476.06 &  0.17\\
 & CDS &  0.07 & 223.24 &  0.52 &  0.34 &  0.12 &  0.32 &  3.21 & 436.81 &  0.18\\
 & SoB & \bf{0.06} & 130.78 &  0.40 & \bf{ 0.31} &  0.10 &  0.28 &  2.97 & 331.27 &  0.14\\
 & CM eigs &  0.08 & 288.41 &  0.55 &  0.60 &  0.15 &  0.53 &  4.54 & 584.44 &  0.21\\
 & Bag of Bonds &  0.07 & 159.24 &  0.47 &  0.35 &  0.12 &  0.28 &  3.34 & 385.59 &  0.18\\
 & \footnotesize{Estate+CDS+SoB} &  0.06 & 129.89 &  0.37 &  0.34 &  0.10 &  0.28 & \bf{ 2.73} & 353.18 &  0.13\\
RF & Estate &  0.09 & 252.74 &  0.59 &  0.50 &  0.14 &  0.39 &  4.09 & 488.98 &  0.19\\
 & CDS &  0.07 & 241.67 &  0.46 &  0.36 &  0.11 &  0.29 &  3.34 & 435.77 &  0.16\\
 & SoB &  0.07 & 136.91 &  0.48 &  0.40 &  0.12 &  0.30 &  3.47 & 417.46 &  0.15\\
 & CM eigs &  0.09 & 286.89 &  0.67 &  0.62 &  0.15 &  0.51 &  5.52 & 512.22 &  0.20\\
 & Bag of Bonds &  0.07 & 172.41 &  0.46 &  0.36 &  0.10 &  0.29 &  3.10 & 418.35 &  0.16\\
 & \footnotesize{Estate+CDS+SoB} &  0.07 & 144.18 &  0.43 &  0.34 & \bf{ 0.09} & 0.26 &  3.11 & 401.27 &  0.15\\
kNN & Estate &  0.08 & 236.55 &  0.61 &  0.49 &  0.15 &  0.41 &  4.30 & 563.89 &  0.20\\
 & CDS &  0.07 & 242.99 &  0.55 &  0.39 &  0.13 &  0.33 &  3.56 & 478.50 &  0.18\\
 & SoB &  0.08 & 184.43 &  0.54 &  0.44 &  0.12 &  0.36 &  3.65 & 427.20 &  0.17\\
 & CM eigs &  0.10 & 343.48 &  0.62 &  0.67 &  0.15 &  0.51 &  5.52 & 570.55 &  0.22\\
 & Bag of Bonds &  0.08 & 238.05 &  0.53 &  0.40 &  0.11 &  0.32 &  3.58 & 515.25 &  0.19\\
 & \footnotesize{Estate+CDS+SoB} &  0.08 & 171.65 &  0.54 &  0.43 &  0.12 &  0.35 &  3.57 & 442.14 &  0.17\\
mean & n/a  & 0.11 & 309.75 &  0.69 &  0.65 &  0.15 &  0.55 &  4.88 & 629.20 &  0.22\\
\hline
     &     &  1.86 &  0.50 &  3.93 &  8.47 &  2.04 &  6.43 &  32.13 &  3568.65 &  1.43 \\
\end{tabular}
\caption{\label{tab:testeverything} Average mean absolute errors (MAEs) in the test sets for different combinations of target property, model and featurization. Hyperparameter optimization was used throughout with nested 5-fold cross validation. The test MAEs are averaged over 20 test sets using shuffle split with 80/20 splitting. The properties are density, heat of formation of the solid, explosive energy, shock velocity, particle velocity, sound velocity, detonation pressure, detonation temperature, and TNT equivalent per cubic centimeter. The models are kernel ridge regression (KRR), ridge regression (Ridge),  support vector regression (SVR), random forest (RF), $k$-nearest neighbors (kNN), and a take-the-mean dummy predictor. The last row gives the average value for each property in the dataset.}
\end{table*}
\subsection{Comparison of machine learning models}

Table \ref{tab:testeverything} presents a comparison of five different ML models \& seven featurization methods for each target property in the Huang \& Massa dataset (a complete comparison with 5 additional models and additional evaluation metrics can be found in Supplementary tables S1, S2, S3, and S4 and in Supplementary Fig. S2). The mean average error was averaged over 20 random train-test splits (with replacement) with a train/test ratio of 4:1. We also calculated 95\% confidence intervals (shown in Supplementary table S) which show the values are well converged and meaningful comparisons can be made. Hyperparameter optimization was performed on all models using grid search. We found that careful hyperparameter optimization, especially of the regularization parameter, is critical to properly comparing and evaluating models. We found that LASSO regression and Bayesian ridge regression performed nearly as well as ridge regression, and gradient boosted trees performed nearly as well as random forest, so we omitted them from the table. Two key observations can be made. The first is that the sum over bonds featurization is the best for all target properties with the exception of the speed of sound, where bag of bonds does slightly better. The second is that kernel ridge and ridge regression performed best. Other models might become competitive with the addition of more data. The gap between the train and test MAEs indicates overfitting is present.

\subsection{Incorporating dimensionality reduction}
Dimensionality reduction can often improve model performance, especially when $n_{\ff{features}} \approx n_{\ff{examples}}$.\cite{Theodoridis2008book} Our first test was done with Coulomb matrices, where we found the error converged with $D=15$ or more principle components. A similar convergence at $D=15$ was observed for E-state and for the combined featurization Estate+SoB+CDS. We also experimented with some other dimensionality reduction techniques such as t-SNE, PCA followed by fast independent component analysis (ICA), and spectral embedding (Laplacian eigenmaps). In all cases, however, the error was not improved by dimensionality reduction (see supplementary information for more detail). This indicates that while our feature vectors could be compressed without loosing accuracy, the regularization in our models is capable of handling the full dimensionality of our feature vectors without loss in performance.

\section{Analysis \& Discussion}

\subsection{Machine learning performance vs. training data quantity \& diversity }

\begin{table}
\centering
\begin{tabular}{ccccc}
 &  & $V_{\ff{det}}$ (km/s)  & $\rho$ (g/cc)  & $P_{\ff{det}}$ (GPa) \\
\hline
KRR& Estate &   0.12, 0.99   &   0.04, 0.98   &   1.99, 0.92  \\
 & CDS & \bf{0.07, 0.99}  &   0.03, 0.99  &  1.10, 0.99  \\
 & SoB &   0.08, 0.99   &   0.03, 0.99   &   \bf{0.83, 0.99}  \\
Ridge & Estate &   0.32, 0.91   &   0.03, 0.98   &   2.48, 0.98  \\
 & CDS &   0.14, 0.99   & \bf{0.02, 0.99}   &   1.34, 0.99  \\
 & SoB &   0.44, 0.86   &   0.03, 0.99   &   2.92, 0.96  \\
mean & n/a &   1.25, 0.00   &   0.27, 0.00   &  12.90, 0.00  \\
\end{tabular}
\caption{Mean absolute errors and Pearson correlation coefficients for ML on the dataset of Ravi et al.\ \cite{Ravi2012:218}, which contains 25 nitropyrazole molecules. 5-fold cross validation was used, so $N_{\ff{train}} = 20$ and $N_{\ff{test}}=5$.}
\label{Ravismall}
\end{table}
To compare our diverse-data case with how machine learning works in a narrow, non-diverse context we fit the dataset of Ravi et al.\ which contains 25 pyrazole-based molecules (table \ref{Ravismall}). Ravi et al.\ performed DFT calculations to obtain the heat of formation with respect to products and then used the Kamlet-Jacobs equations to predict detonation velocity and detonation pressure. Table \ref{Ravismall} shows how very high accuracy in predicting energetic properties is achieved. Not surprisingly, the custom descriptor set slightly outperforms other featurizations here, since the only thing that is changing within the dataset is the type and number of functional groups attached to the pyrazole backbone. While accurate, the trained models will not generalize beyond the class of molecules they were trained since they cannot capture significant differences between classes.\cite{Zeman2007Book} Similarly, insights gained from the interpretation of such models may not generalize and may be contaminated by spurious correlations that often occur in small datasets.

The part of chemical space where a model yields accurate predictions is known as the applicability domain.\cite{Sahigara2012:4791} The concept of applicability domain is inherently subjective as it depends on a choice of accuracy one considers acceptable. Mapping the applicability domain of a model by generating additional test data is typically expensive given the high dimensionality of chemical space. There are several heuristics for estimating an outer limit to the applicability domain of a model, such as using the convex hull of the training data, a principle components based bounding box, or a cutoff based on statistical leverage.\cite{Sahigara2012:4791} While useful, applicability domains determined from such methods all suffer from the possibility that there may be holes inside where the model performs badly. To spot holes, distance based methods can be used, which average the distance to the $k$ nearest neighbors in the training data under a given metric and apply a cutoff. There is no universal way of determining a good cutoff, however, so determining a practical cutoff requires trial and error.


\subsection{Residual analysis }
\begin{figure*}[ht]
 \includegraphics[width=18cm]{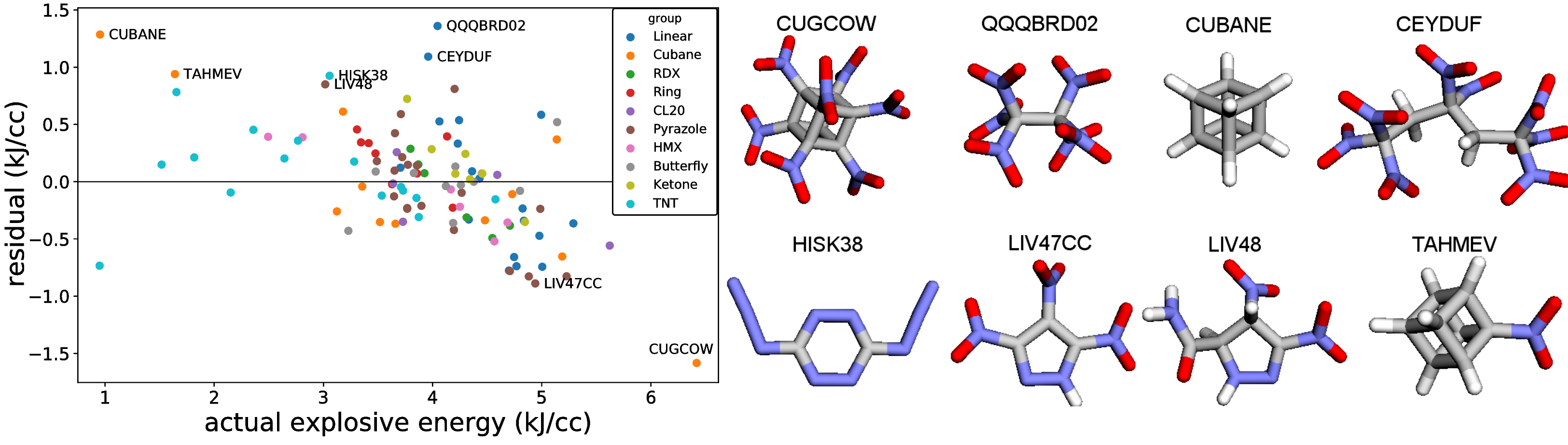}
  \caption{\label{EEresiduals} Residuals in leave-one-out cross validation with kernel ridge regression and the sum over bonds featurization (left), and some of the worst performing molecules (right).}
\end{figure*}
\begin{table}[ht]
\centering
\begin{tabular}{c c c c c c}
     group   & N$_{\ff{group}}$ & $r$ & R$^2$ & MAE  & avg.\ residual (kJ/cc) \\
\hline
                          HMX  &   6 &  0.968 &   0.83  &   0.32 &  -0.06   \\
                    Butterfly  &  10 &  0.917 &   0.78  &   0.18 &  -0.01   \\
                          TNT  &  16 &  0.854 &   0.83  &   0.31 &   0.10   \\
                         CL20  &   6 &  0.854 &   0.83  &   0.22 &  -0.09   \\
                       Cubane  &  12 &  0.814 &   0.75  &   0.58 &  -0.04   \\
                         Ring  &   8 &  0.548 &   0.17  &   0.26 &   0.20   \\
                          RDX  &   6 &  0.377 &   0.19  &   0.28 &  -0.11   \\
                     Pyrazole  &  20 &  0.254 &   0.21  &   0.42 &  -0.07   \\
                       Ketone  &   7 &  0.099 &  -0.13  &   0.25 &   0.15   \\
                       Linear  &  18 &  0.003 &  -1.12  &   0.52 &   0.00   \\
\end{tabular}
  \caption{The mean absolute error, Pearson correlation and average residual in different groups, for prediction of explosive energy (kJ/cc) with leave-one-out CV on the entire dataset using sum over bonds and kernel ridge regression. The groups are sorted by $r$ value rather than MAE since the average explosive energy differs between groups.}
  \label{groupresiduals}
\end{table}

Useful insights into the applicability of the model can be gained by looking at the residuals ($y_{\ff{pred}} - y_{\ff{true}}$) in the test set. We chose to look at residuals in leave-one-out cross validation using the sum over bonds featurization and kernel ridge regression (fig.\ \ref{EEresiduals}). We suspected that our model may have difficulty predicting the explosive energy for cubane-derived molecules, since they release a large amount of strain energy in explosion - something which is not significant in other molecules except for the CL20 group, and not explicitly contained in any of our featurizations. In fact, the three worst performing molecules were heptanitrocubane, the linear molecule QQQBRD02, and cubane. The model overestimates the energy of cubane and underestimates the explosive energy of nitrocubane. The mean absolute errors and average residuals of different groups in leave-one-out CV are shown in table \ref{groupresiduals}.
In order to visualize the distribution of data in high dimensional space one can utilize various embedding techniques that can embed high dimensional data into two dimensions. Often one may find that high dimensional data lies close to a lower dimensional manifold. The embeddings shown in the supplementary information (t-SNE, PCA, \& spectral) show that the cubane molecules are quite separated from the rest of the data (Supplementary Information Fig. S3). Based on this fact, we tried removing the cubanes. We found that the accuracy of prediction of explosive energy with kernel ridge regression and sum over bonds remained constant (MAE = 0.36 kJ/cc) while the $r$ value actually decreased significantly from $0.76$ to $0.68$.


\subsection{Learning curves}

\begin{figure*}[ht]
 \centering
    \includegraphics[height=6cm]{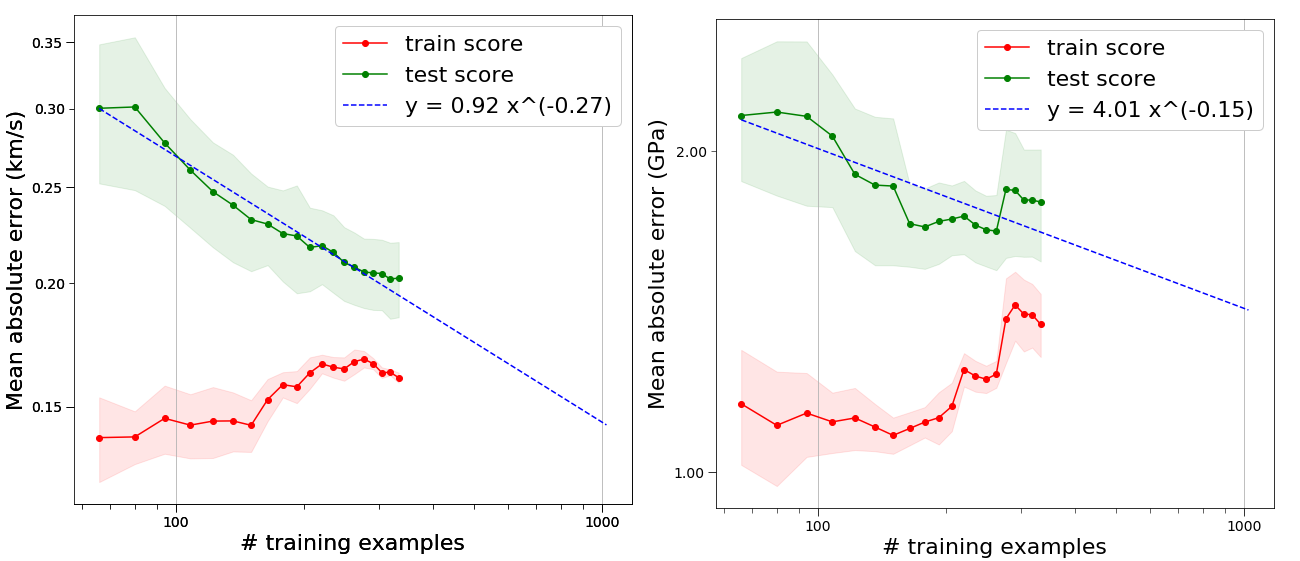}
    \caption{The learning curves for predicting detonation velocity (left) and detonation pressure (right) for the combined ($N = 418$) dataset plotted on a log-log plot. Shaded regions show the standard deviation of the error in 5-fold cross validation.}
     \label{learningcurve}
\end{figure*}

Insights into the data-dependence of ML modeling can be obtained from plots of cross-validated test error vs number of training examples, which are known as learning curves. While we were able to obtain good learning curves from just the Huang \& Massa dataset, to ensure their accuracy we supplemented the Huang \& Massa data with 309 additional molecules from the dataset given in the supplementary information of Mathieu et al., which includes detonation velocity and detonation pressure values calculated from the Kamlet-Jacobs equations.\cite{Mathieu2017:8191} A few of the molecules are found in both datasets, but most are unique, yielding a total of $\approx 400$ unique molecules. We assume detonation velocity is equivalent to shock velocity in Huang \& Massa data. The method of calculation of properties differs between the two datasets, possibly introducing differing bias between the sets, so we shuffle the data beforehand. Figure \ref{learningcurve} shows the learning curves for detonation velocity and detonation pressure. The gap between training and test score curves indicates error from overfitting (variance) while the height of the curves indicates the degree of error from the choice of model (bias). As the quantity of data increases, the gap between the training and test curves should decrease. The flattening out of the training accuracy curve indicates a floor in accuracy for the given choice of model \& featurization. Even with much more data, it is very likely that creating predictions more accurate than such a floor would require better choices of model and featurization. Empirically, learning curves are known to have a $A N^{-\beta}_{\ff{train}}$ dependence.\cite{Huang2016:161102} In the training of neural networks typically $1 < \beta < 2$,\cite{Muller1996:1085} while we found values between $0.15-0.30$ for the properties studied. Since different types of models can have different learning curves, we also looked at random forest, where we found similar plots with $\beta \approx 0.20$.

\section{Conclusion \& Future Directions}
We have presented evidence that machine learning can be used to predict energetic properties out of sample after being trained on a small yet diverse set of training examples ($N_{\ff{train}} = 87$, $N_{\ff{test}}= 22$). For all the properties tested, either kernel ridge or ridge regression were found to have the highest accuracy, and out of the base featurizations we compared the sum over bonds featurization always performed best. Small improvements could sometimes be gleaned by concatenating featurizations. The best $r$ values achieved were 0.94 for heat of formation, 0.74 for density, 0.79 for explosive energy, and 0.78 for shock velocity. With kernel ridge regression and sum over bonds we obtained mean percentage errors of 11\% for explosive energy, 4\% for density, 4\% for detonation velocity and 11\% for detonation pressure. By including $\approx 300$ additional molecules in our training we showed how the mean absolute errors can be pushed lower, although the convergence with number of molecules is slow.

There are many possible future avenues of research in the application of machine learning methods to the discovery of new energetic materials. One possible way to overcome limitations of small data is with transfer learning.\cite{Hutchinson2017arxiv,Junshui2015:263} In transfer learning, a model is first trained for a task where large amounts of data is available, and then the model's internal representation serves as a starting point for a prediction task with small data. Another way our modeling might be improved without the need for more data is by including in the training set CNOHF type molecules that are non-energetic, which may require overcoming sampling bias by reweighting the energetic molecules. Additional work we have undertaken investigates how our models can be interpreted to illuminate structure-property relationships which may be useful for guiding the design of new energetic molecules.\cite{BarnesIDS2018} Finally, a promising future direction of research involves coupling the property predicting models developed here with generative models such as variational autoencoders or generative adversarial networks to allow for molecular generation and optimization.


\section{Methods}
We used in-house Python code for featurization and the {\it scikit-learn} package (\href{http://scikit-learn.org/}{http://scikit-learn.org/}) for machine learning. Parts of our code has been used to establish a library we call the Molecular Machine Learning (MML) toolkit. The MML toolkit is open source and available online at \href{https://github.com/delton137/mmltoolkit}{https://github.com/delton137/mmltoolkit}.

\subsection{Model evaluation}

There are many different ways to evaluate model performance. The simplest scoring function is the mean absolute error (MAE):
\begin{equation}
    \hbox{MAE} = \frac{1}{N} \sum\limits_i^N \left| y^{\ff{true}}_i - y^{\ff{pred}}_i  \right|
\end{equation}

A related scoring function is the root mean squared error (RMSE), also called the standard error of prediction (SEP), which is more sensitive to outliers than MAE:
\begin{equation}
    \hbox{RMSE} = \sqrt{\frac{1}{N} \sum_i (y^{\ff{true}}_i - y^{\ff{pred}}_i )^2  }
\end{equation}

We also report the mean average percent error (MAPE), although this score is often misleadingly inflated when $y^{\ff{true}}_i$ is small:
\begin{equation}
    \hbox{MAPE} = \frac{100}{N}\sum\limits_i^N \left| \frac{y^{\ff{true}}_i - y^{\ff{pred}}_i   }{y^{\ff{true}}_i } \right|
\end{equation}

It is also important to look at the Pearson correlation coefficient $r$:
\begin{equation}
    r =  \frac{\sum_i (y^{\ff{true}}_i - \bar{y}^{\ff{true}}_i)(y^{\ff{pred}}_i - \bar{y}^{\ff{pred}}_i)
    }{
    \sqrt{ \sum_i (y^{\ff{true}}_i - \bar{y}^{\ff{true}}_i)^2 \sum_i (y^{\ff{pred}}_i - \bar{y}^{\ff{pred}}_i)^2 } }
\end{equation}

Additionally, we report the coefficient of determination:
\begin{equation}\label{coeffdeterm}
    R^2 = 1 - \frac{\sum_i (y^{\ff{true}}_i - y^{\ff{pred}}_i)^2   }{ \sum_i (y^{\ff{true}}_i - \bar{y}^{\ff{true}}_i)^2  }
\end{equation}
Unlike the Pearson correlation coefficient $r$, $R^2$ can assume any value between -$\infty$ and 1. When reported for test or validation data, this scoring function is often called $Q^2$. While a bad $Q^2$ indicates a bad model, a good $Q^2$ does not necessarily indicate a good model, and models should not be evaluated using $Q^2$ in isolation.\cite{Golbraikh2002:269}


\subsection{Data gathering}
With a the exception of the Coulomb-matrix and bag-of-bonds featurizations, the only input required for our machine learning featurizations is the molecular connectivity graph. For easy encoding of the molecular graph we used Simplified Molecular-Input Line-Entry System (SMILES) strings.\cite{Weininger1988} SMILES strings are a non-unique representation which encode the molecular graph into a string of ASCII characters. SMILES strings for 56 of the Huang \& Massa molecules were obtained from the Cambridge Structure Database Python API,\cite{Groom2016:171} and the rest were generated by hand using a combination of the Optical Structure Recognition Application,\cite{Filippov2009:740} the \href{www.molview.org}{www.molview.org} molecule builder, and the Open Babel package\cite{OBoyle2011:33}, which can convert .mol files to SMILES. Since Coulomb matrices require atomic coordinates, 3D coordinates for the molecules were generated using 2D$\rightarrow$3D structure generation routines in the RDKit\cite{rdkit} and Open Babel\cite{OBoyle2011:33} python packages. After importing the SMILES into RDKit and adding hydrogens to them, an embedding into three dimensions was done using distance geometry, followed by a conjugate gradient energy optimization. Next, a weighted rotor search with short conjugate gradient minimizations was performed using Open Babel to find the lowest energy conformer. Finally, a longer conjugate gradient optimization was done on the lowest energy conformer. For one molecule (the cubane variant `EAT07') we used the {\it obgen} utility program of Open Babel to do the coordinate embedding (in retrospect, {\it obgen} could have been used to yield good enough results for all molecules). All energy minimizations were done with the MMFF94 forcefield.\cite{Halgren1996:490} The accuracy of the generated structures was verified by visually comparing the generated structures of 56 of the compounds to x-ray crystallographic coordinates obtained from the Cambridge Structure Database.

\subsection{Data Availability}
The compiled property data from Huang \& Massa (2010) and Mathieu (2017) and the generated SMILES strings are included in the Supplementary Information. The full Mathieu (2017) dataset is available free of charge on the ACS Publications website at \href{http://pubs.acs.org/doi/suppl/10.1021/acs.iecr.7b02021}{DOI: 10.1021/acs.iecr.7b02021} Python Jupyter notebooks for reproducing all of our results have been open sourced and are available at \href{https://github.com/delton137/Machine-Learning-Energetic-Molecules-Notebooks}{https://github.com/delton137/Machine-Learning-Energetic-Molecules-Notebooks}. The Python notebooks use the open source Molecular Machine Learning toolkit developed in conjunction with this work, which can be found at \href{https://github.com/delton137/mmltoolkit}{https://github.com/delton137/mmltoolkit}.

\bibliography{Machine_learning_energetics}

\section*{Acknowledgements}
Support for this work is gratefully acknowledged from the U.S. Office of Naval Research under grant number N00014-17-1-2108 and from the Energetics Technology Center under project number 2044-001. Partial support is also acknowledged from the Center for Engineering Concepts Development in the Department of Mechanical Engineering at the University of Maryland, College Park. We thank Dr. Brian C. Barnes for comments on the manuscript. We also thank Dr. Ruth M. Doherty and Dr. William Wilson from the Energetics Technology Center for their encouragement, thoughts on energetic materials, and for proofreading the manuscript.

\section*{Author contributions statement}
P.W.C. and M.F. conceived the research, D.E. wrote the code and most of the manuscript, M.B. prepared the data and helped generate the SMILES strings, Z.B. did part of the machine learning and dimensionality reduction work. All authors reviewed and edited the manuscript.

\section*{Additional information}
The author(s) declare no competing interests.




\end{document}